\documentclass[amsmath,nofootinbib,notitlepage,prd,superscriptaddress,twocolumn]{revtex4-2}

\usepackage[T1]{fontenc}
\usepackage[usenames,dvipsnames]{xcolor}
\usepackage{aas_macros,graphicx,hyperref,orcidlink}
\usepackage{amsfonts}

\definecolor{mycolor}{RGB}{153, 0, 0}
\hypersetup{
    colorlinks=true,
    linkcolor=mycolor, 
    citecolor=mycolor, 
    urlcolor=mycolor  
 }

\newcommand{\br}[1]{\left[#1\right]}

\newcommand{\pa}[1]{\left(#1\right)}
\newcommand{\ed}{\mathop{}\!\mathrm{d}}

\newcommand{\HH}{\mathrm{H}}

\newcommand{\tot}{\mathrm{tot}}
\newcommand{\TT}{\mathrm{TT}}

\newcommand{\fr}{\epsilon_\mathrm{frac}}
\newcommand{\Nharm}{N_\mathrm{harm}}

\newcommand{\atan}[1]{\mathrm{arctan}\left(#1\right)}

\usepackage[T1]{fontenc}
\begin{document}

\title{Estimating High-Order Time Derivatives of Kerr Orbital Functionals}

\author{Lennox S. Keeble\,\orcidlink{0009-0009-5796-631}} 
\email{lkeeble@princeton.edu}
\affiliation{Department of Physics, Princeton University, Princeton, New Jersey 08544, USA}

\author{Alejandro C\'ardenas-Avenda\~no\,\orcidlink{0000-0001-9528-1826}} 
\email{cardenas@wfu.edu}
\affiliation{Department of Physics, Wake Forest University, Winston-Salem, North Carolina 27109, USA}

\begin{abstract}
Functions of bound Kerr geodesic motion play a central role in many calculations in relativistic astrophysics, ranging from gravitational-wave generation to self-force and radiation-reaction modeling. Although these functions can be expressed as a Fourier series using the geodesic fundamental frequencies, reconstructing them in coordinate time is challenging due to the coupling of the radial and polar motions. In this paper, we compare two strategies for performing such reconstructions and their ability to estimate high-order coordinate-time derivatives of the orbital functional. The first method maps Fourier coefficients from Mino to coordinate time; the second method fits a sampled time series of the function to a truncated coordinate‑time Fourier series. While the latter method is prone to overfitting, it yields more accurate reconstructions and derivatives than the mapping, but completely misrepresents the harmonic content of the orbital functional. For the purpose of accurate coordinate-time derivative estimation, we propose a hybrid method: fit for the Mino‑time coefficients, differentiate with respect to Mino time, then convert to coordinate time. Applied to the mass quadrupole of a generic Kerr geodesic, this hybrid method recovers the sixth derivative with a fractional residual $\sim10^{-6}$ using only two harmonics. For orbital functionals that depend explicitly on the geodesic orbit expressed in Boyer--Lindquist coordinates, we also provide a recursive procedure for computing coordinate-time derivatives using exact analytic expressions. These results offer a general framework for accurately evaluating high-order time derivatives along Kerr geodesic worldlines, with direct relevance to applications such as extreme-mass-ratio inspiral kludge waveform modeling, where such derivatives are key ingredients for precise gravitational-wave predictions.

\end{abstract}

\maketitle
\section{Introduction}
\label{sec:Introduction}

Evaluating time derivatives along a Kerr orbit’s worldline is central to relativistic orbital dynamics and gravitational-wave physics. Radiation formulas, from the quadrupole approximation to higher-order multipole expansions, require these derivatives of the source’s multipole moments~\cite{Blanchet:1995fg}. Time derivatives of the orbit yield the changing mass quadrupole and higher moments, which are crucial for determining instantaneous power emission and energy-angular momentum fluxes.

These high-order time derivatives arise in many settings, from binary inspirals to scattering encounters, and, more generally, wherever radiation reaction or wave emission is modeled. In post-Newtonian and post-Minkowskian approaches, bodies are described by worldlines endowed with multipole moments, whose time derivatives are essential for accurate flux calculations, particularly when spin or tidal effects are included~\cite{Kidder:1992fr,Blanchet_2010,Henry:2019xhg}.

High-order time derivatives of Kerr orbits also play an especially important role in one of the most important astrophysical applications of black hole perturbation theory: modeling the spacetime perturbations of a body orbiting around a massive compact object and the radiative degrees of freedom which determine the gravitational wave signal a distant observer measures~\cite{Poisson:2011nh,Pound_2021}. Examples of such astrophysical systems are extreme-mass-ratio inspirals (EMRIs), which, in the typical picture, consist of a stellar mass black hole inspiraling into a supermassive black hole at the center of a galaxy. EMRI systems are expected to be key sources for low-frequency space-based gravitational wave observatories, such as the Laser Interferometer Space Antenna~\cite{Amaro-Seoane_2017,Babak_2017,Colpi_2024,LISA:2022kgy}, and are expected to provide detailed information of the spacetime~\cite{Vílchez_2025, Speri_2024, Katz_2020, Katz_2021, Marsat_2021}, therefore providing an exceptional laboratory for testing general relativity in the strong-field regime~\cite{Barack_2019, Cardenas_Avendano_2024}.

The application of black hole perturbation theory to EMRI modeling typically proceeds by expanding the perturbed metric in powers of the mass-ratio parameter $q=m/M\ll1$, where $M$ and $m$ denote, respectively, the mass of the primary and secondary, and considering the perturbative degrees of freedom order by order. First-order perturbations are governed by the Teukolsky equation, with the radiative degrees of freedom encapsulated in the two Weyl scalars $\psi_{0}$ and $\psi_{4}$~\cite{Teukolsky_1973, Wald_1973}. To date, frameworks based on the Teukolsky equation for generating EMRI inspirals and waveforms in the adiabatic limit—where the first-order gravitational self-force is averaged over a slowly-evolving non-resonant orbit—have made significant advancement~\cite{Hughes_2021}. Post-adiabatic effects such as resonant orbits~\cite{Flanagan_2012, Flanagan_2014, Ruangsri_2014, Speri_2021, Levati_2025}, spin-curvature coupling~\cite{Drummond_1_2022, Drummond_2_2022, Skoupy_2023, Drummond_2023, Drummond_2024}, the conservative part of the self-force~\cite{babak2007kludge} and second-order effects~\cite{Pound_2020, Pound_2021} are expected to be crucial in modeling EMRIs with the level of accuracy that will be necessary for next-generation space-based detectors~\cite{LISAConsortiumWaveformWorkingGroup:2023arg}.

From so-called ``kludge'' schemes—where one combines post-Newtonian, post-Minkowskian, and any other black hole perturbation technique to quickly generate approximate EMRI waveforms~\cite{Barack_2004, babak2007kludge, Sopuerta_2011, Chua_2017,Katz:2021yft,Chapman-Bird:2025xtd}—to more robust but computationally expensive approaches, such as the Teukolsky formalism~\cite{Hughes_2021} or a detailed self-force prescription~\cite{Pound_2021,Nasipak_2025,Lewis_2025}, it is often essential to utilize the symmetry properties of the secondary body's leading-order geodesic motion in the Kerr background of the massive object. Of particular use is the Fourier-domain representation of Kerr geodesic orbits and functions thereof (or ``orbital functionals''), whose existence is guaranteed due to geodesic motion in Kerr spacetime possessing four constants of motion, as can be shown using Hamilton--Jacobi techniques~\cite{Carter_1968} or the action-angle variable formalism~\cite{Schmidt_2002}. Some approximate waveform models require high-order multipole moments and self-force terms which involve high-order time derivatives of orbital functionals. While computing these derivatives reliably is not a trivial task numerically or otherwise, the Fourier-domain representation provides a particularly elegant framework for approximating them~\cite{Sopuerta_2011}. 

An orbital functional $f(\textsf{T})\!=\!f[r(\textsf{T}),\theta(\textsf{T}),\phi(\textsf{T})]$, where $\textsf{T}$ is some time parameter, possesses a Fourier expansion in terms of harmonics of the fundamental frequencies of motion with respect to $\textsf{T}$~\cite{Schmidt_2002}. With some prescription for determining the fundamental frequencies and the Fourier coefficients in this expansion, one can then, in principle, reconstruct $f(\textsf{T})$. With such a reconstruction, one can estimate time derivatives of $f(\textsf{T})$ using the property that the time derivative operator becomes a simple algebraic operator in the Fourier domain, with each term in the Fourier expansion being multiplied by its corresponding harmonic frequency~\cite{Sopuerta_2011}.

In astrophysical applications, it is often necessary to derive observables in terms of some set of orbital functionals, utilizing a time parameter that aligns with the rate at which the clocks of distant observers tick. As addressed in Ref.~\cite{Drasco_2004}, however, the coupled nature of the radial and polar Kerr geodesic motion with respect to coordinate time (e.g., Boyer--Lindquist time $t$) makes the computation of the $t$-expansion Fourier coefficients challenging. However, the Mino time parameter $\lambda$ is especially useful because the radial and polar motion decouple with respect to $\lambda$~\cite{Mino_2003}. This decoupling was utilized in Ref.~\cite{Drasco_2004} to prescribe a method for extracting the $t$-expansion Fourier coefficients from the more accessible $\lambda$-expansion coefficients. In principle, these expansion coefficients should also be attainable by fitting a sampled time series array of $f(t)$ to a truncated version of its Fourier expansion for the coefficients, as proposed in Ref.~\cite{Sopuerta_2011}. 

In this work, building on the Fourier‑coefficient mapping of Ref.~\cite{Drasco_2004} (first method) and the time‑series fitting strategy of Ref.~\cite{Sopuerta_2011} (second method), we systematically compare how well the second method recovers the Fourier coefficients as given fiducially by the first, as well as the respective accuracy of each resulting derivative estimation. Focusing on derivatives up to sixth order—readily extendable to higher orders—we benchmark accuracy and efficiency for an orbital functional with a simple analytic form and a set of more complex functions. We also introduce a refinement of the fitting method proposed in Ref.~\cite{Sopuerta_2011} which first determines the $\lambda$‑expansion coefficients of $f(\lambda)$, computes derivatives with respect to Mino time, and then converts these to coordinate time. In the examples considered in this work, we find this modified approach to be the most accurate for high-order time derivative estimation.

As applied to the (seemingly) simple orbital functional $f(t)=r(t)\cos\theta(t)$, also considered in Ref.~\cite{Drasco_2004}, we find that the second method tends to overfit the time series. On one hand, this overfitting results in a failure to accurately capture the ``true'' power spectrum as obtained from the first method, but, on the other hand, it provides time series reconstructions and derivative estimations which are orders of magnitude more accurate than the first method for a given truncation of the Fourier expansion. We illustrate a similarly accurate and efficient application of the fitting method to estimating time derivatives of the mass and current multipoles used in the kludge scheme introduced in Ref.~\cite{Sopuerta_2011}. In both examples, the refined fit approach using the $\lambda$-expansion is approximately two orders of magnitude more accurate than the fitting method implemented in Ref.~\cite{Sopuerta_2011}.

We also describe a systematic procedure for computing high-order coordinate time derivatives of geodesic orbits directly from the equations of motion, which is necessary for setting the benchmark when assessing the accuracy of the two approximation methods. As discussed in Ref.~\cite{Sopuerta_2011}, deriving analytic expressions for these time derivatives which can be evaluated efficiently and accurately is nontrivial due to the complex expressions that arise from differentiating the equations of motion without careful consideration. In this work, we sidestep this difficulty by leveraging the separated nature of the $r$ and $\theta$ first-order geodesic equations with respect to Mino time. In particular, we describe a systematic procedure in which one first computes derivatives with respect to Mino time and then transforms these to coordinate time derivatives. This procedure is also necessary for implementing our modified approach of the fitting scheme, and, as we illustrate in Sec.~\ref{ssec:MultipoleExample}, it can additionally be used to compute coordinate time derivatives of orbital functionals which depend explicitly on the orbit (but which do not necessarily need a closed-form expression in terms of $r$, $\theta$, and $\phi$).

We emphasize that this work is concerned with the pointwise accuracy of the derivative estimation methods we investigate. We assume that one is interested in estimating a time derivative at some point along a geodesic trajectory that lies in the center of an array. This can generally be achieved because of the time-reversibility of the Kerr solution, which allows one to evolve a geodesic into the past and future of a given point in spacetime to artificially place that point at the center of a time series array. Throughout, we therefore focus on the fractional residual near the center of the time series arrays and not on the global nature of the errors, which can generally be much larger at the endpoints.

Since we exploit known properties of geodesics in the Kerr background in this work, in what follows, we only provide the main elements essential to understanding our work and refer the reader to the seminal papers we cite. We also do not report explicit run times for our numerical implementation of the methods discussed in this paper since our implementations have not gone through any optimization processes.

The remainder of this paper is organized as follows. We begin by briefly reviewing in Sec.~\ref{sec:FourierReconstruction} the salient features of reconstructing time series of orbital functionals from their Fourier series expansion in the fundamental frequencies of motion. In Sec.~\ref{sec:BLDerivs}, we outline an efficient procedure for computing high-order coordinate time derivatives of geodesic orbits which are necessary for the modified derivative estimation approach we suggest. Then, in Sec.~\ref{sec:Examples}, we present two examples of derivative estimation of orbital functionals. In Sec.~\ref{ssec:SimpleExample}, we compare the power spectrum and accuracy of the two derivative estimation methods applied to a simple orbital functional. In Sec.~\ref{ssec:MultipoleExample}, we consider the more complex orbital functionals of mass and current multipoles expressed in harmonic coordinates. We conclude in Sec.~\ref{sec:Discussion} with a summary and discussion of our results. We work in geometric units where $G=1=c$ and with the $(-,+,+,+)$ metric signature.

\section{Fourier series reconstruction of orbital functionals}
\label{sec:FourierReconstruction}

In this section, we review the key features of reconstructing time series of orbital functionals from their Fourier series expansion in the fundamental frequencies of motion.

Bound, timelike geodesic motion of a test particle in Kerr spacetime is governed by the coupled set of first-order ordinary differential equations (ODEs)~\cite{Carter_1968,Chandrasekhar_1983}
\begin{subequations}
\begin{align}
    \Sigma(r, \theta)\frac{\ed t}{\ed\tau}&= T(r, \theta),\label{eq:PropTimeODEt}\\
    \Sigma(r, \theta)^2\left(\frac{\ed r}{\ed\tau}\right)^2&=R(r),\label{eq:PropTimeODEr}\\
    \Sigma(r, \theta)^2\left(\frac{\ed\theta}{\ed \tau}\right)^2&=\Theta(\theta),\label{eq:PropTimeODEθ}\\
    \Sigma(r, \theta)\frac{\ed\phi}{\ed\tau}&= \Phi(r, \theta),\label{eq:PropTimeODEphi}
\end{align}\label{eq:PropTimeGeoODEs}
\end{subequations}
where $\tau$ denotes the proper time, $\Sigma\equiv r^{2}+a^{2}\cos^{2}{\theta}$, $\Delta\equiv r^{2}-2Mr+a^{2}$, and
\begin{subequations}
    \begin{align} R\pa{r}&=\br{\pa{r^{2}+a^{2}}\,E-aL_{z}}^{2}\nonumber\\
    &\quad-\Delta\br{r^{2}+\pa{L_{z}-aE}^{2}+Q},\label{eq:Rfunc}\\
    \Theta\pa{\theta}&=Q-a^{2}(1-E^{2})\cos^{2}{\theta}-L_{z}^{2}\cot^{2}{\theta} ,\label{eq:ThetaFunc}\\
        T(r, \theta)&\equiv E\br{\frac{\pa{r^{2}+a^{2}}^{2}}{\Delta}-a^{2}\sin^{2}{\theta}}\nonumber\\
        &\quad+aL_{z}\br{1-\frac{r^{2}+a^{2}}{\Delta}},\label{eq:TFunc}\\
        \Phi(r, \theta)&\equiv\csc^{2}{\theta}L_{z}+aE\pa{\frac{r^{2}+a^{2}}{\Delta}-1}-\frac{a^{2}L_{z}}{\Delta}.
    \end{align}
\end{subequations}

As first shown by Carter~\cite{Carter_1968}, the system of first-order ODEs (Eq.~\ref{eq:PropTimeGeoODEs}) can be derived from Hamilton--Jacobi theory and arise from the existence of four constants of geodesic motion in Kerr: the rest mass $m$ of the test particle and the constants $E$, $L_{z}$, and $Q$, which, respectively, correspond to the specific energy, $z$-component of the angular momentum and Carter constant.

One can write the geodesic equations (Eq.~\ref{eq:PropTimeGeoODEs}) with respect to coordinate time, $t$, the rate at which the clocks of distant observers tick, by dividing Eqs.~(\ref{eq:PropTimeODEr}--\ref{eq:PropTimeODEphi}) by Eq.~\ref{eq:PropTimeODEt}. However, with either $t$ or $\tau$ as the time parametrization, the ODEs governing the radial and polar motion remain coupled. Following its use by Mino~\cite{Mino_2003} in parametrizing separated radial and polar motions of Kerr geodesic orbits, the Mino time parameter $\lambda$, defined by $\ed\lambda = \ed \tau/ \Sigma$, has proven particularly useful in studying strong-field black hole orbits. With $\lambda$ as the time parameter, the geodesic equations (Eq.~\ref{eq:PropTimeGeoODEs}) become
\begin{subequations}
\begin{align}
    \frac{\ed t}{\ed\lambda}&= T(r, \theta),\label{eq:MinoTimeODEt}\\
    \left(\frac{\ed r}{\ed\lambda}\right)^2&=R(r),\label{eq:MinoTimeODEr}\\
    \left(\frac{\ed\theta}{\ed\lambda}\right)^2&=\Theta(\theta),\label{eq:MinoTimeODEθ}\\
    \frac{\ed\phi}{\ed\lambda}&= \Phi(r, \theta).\label{eq:MinoTimeODEphi}
\end{align}\label{eq:MinoTimeGeoODEs}
\end{subequations}As is evident from Eqs.~(\ref{eq:MinoTimeODEr}--\ref{eq:MinoTimeODEθ}), the radial and polar motion each completely decouple from the rest of the system. A consequence of this decoupling is that $r$ and $\theta$ are periodic functions of $\lambda$, meaning that there exists well-defined periods $\Lambda_{r},\Lambda_{\theta}$ such that $r(\lambda)=r(\lambda+n\Lambda_{r})$, with $n \in \mathbb{Z}$ and likewise for $\theta$---no such well-defined periods exist with respect to coordinate time. As explicitly demonstrated in Ref.~\cite{Drasco_2004} and discussed below, the separation of the radial and polar motion in Mino time can dramatically simplify calculations which would otherwise be intractable in coordinate time.

Through the use of the action-angle variable formalism, one can show that generic orbital functionals possess a coordinate time expansion~\cite{Goldstein, Schmidt_2002,Drasco_2004},
\begin{align}
    f\left[r(t), \theta(t), \phi(t)\right]=\sum_{m, k,  n}f_{m,k,n}\,e^{-i\Omega_{m,k,n}t},\label{eq:Kerr:BLExpansion}
\end{align}
where each integer index in the sum runs from $-\infty$ to $\infty$, $f(r, \theta, \phi)$ being real requires $f_{-m,-k,-n}=\bar{f}_{m,k,n}$ (with bar notation denoting the complex conjugate), and we have defined $\Omega_{m,k,n}$ in terms of the three fundamental frequencies of motion with respect to coordinate time,
\begin{align}
    \Omega_{m,k,n}&=m\Omega_{\phi}+k\Omega_{\theta}+n\Omega_{r}.
\end{align}
The Fourier coefficients in Eq.~\ref{eq:Kerr:BLExpansion} are given by
\begin{align}
    f_{m,k,n}=\frac{1}{(2\pi)^{3}} \iiint _{0}^{2\pi}  f(r, \theta, \phi)e^{i\left(mw^{\phi}+kw^{\theta}+nw^{r}\right)}\ed\vec{w},\label{eq:Kerr:FCoeffs}
\end{align}
where $\ed\vec{w}=\ed w^{\phi}\ed w^{\theta}\ed w^{r}$ and the $w^{i}=\Omega^{i}t$ are angle variables defined with respect to coordinate time. We can also write such an expansion in Mino time~\cite{Drasco_2004}
\begin{align}
    f\left[r(\lambda), \theta(\lambda), \phi(\lambda)\right]=\sum_{m, k, n}f_{m,k,n}\,e^{-i\Upsilon_{m,k,n}\lambda},\label{eq:Kerr:MinoExpansion}
\end{align}where the Fourier coefficients are also given by Eq.~\ref{eq:Kerr:FCoeffs}, except now the angle variables are with respect to Mino time, i.e., $w^{i}=\Upsilon^{i}\lambda$, where the $\Upsilon^{i}$ denote the radial, polar and azimuthal fundamental frequencies with respect to Mino time. If the orbital functional depends only on $r$ and $\theta$, the expansions in Eqs.~(\ref{eq:Kerr:BLExpansion}, \ref{eq:Kerr:MinoExpansion}) will only have nonzero coefficients $f_{0,k,n}\equiv f_{k,n}$.

In astrophysical applications, the expansion in coordinate time is more useful, but the extraction of its Fourier coefficients is generally more difficult than in Mino time~\cite{Drasco_2004}. In order to obtain the Fourier coefficients from Eq.~\ref{eq:Kerr:FCoeffs}, one must be able to independently evaluate $f(r, \theta, \phi)$ as a function of either the coordinate time or Mino time angle variables. The difficulty which presents itself in coordinate time is that the radial and polar motions are coupled. Consequently, both $r$ and $\theta$ are functions of each other's angle variables, e.g., $r=r(w^{r}, w^{\theta})$, so that, for example, reconstructing the simple function $f(r, \theta, \phi) = r(t)$ from its Fourier series expansion requires the evaluation of the following integral:
\begin{align}
    f_{k,n}=\frac{1}{(2\pi)^{2}} \iint_{0}^{2\pi} r(w^{r}, w^{\theta})e^{i\left(kw^{\theta}+mw^{r}\right)}\ed w^{\theta}\ed w^{r}.\label{eq:BLrReconstruction}
\end{align}
One typically solves the first-order geodesic equations numerically for the trajectory, the result of which is a time series array $r(t)$. Equation \ref{eq:BLrReconstruction}, however, cannot be evaluated using this time series since one must specify two independent values of $w^{r}=\Omega^{r}t$ and $w^{\theta}=\Omega^{\theta}t$ to evaluate $r(w^{r}, w^{\theta})$, but these values of the angle variables will generally map onto two distinct values of $t$. Conversely, in Mino time, the radial and polar motions separate so that $r=r(w^{r})$ has an expansion only in the angle variable $w^{r}$ and one is able to evaluate $r(w^{r})=r(\lambda=w^{r}/\Upsilon^{r})$ using interpolation, allowing for the computation of the Fourier coefficients $f_{k}$ from a one-dimensional version of Eq.~\ref{eq:BLrReconstruction}, and thus one can straightforwardly reconstruct the time series $r(\lambda)$.

In the following sections, we compare two methods for calculating the coordinate time Fourier coefficients. The difficulty of extracting the coordinate time Fourier coefficients was addressed in Ref.~\cite{Drasco_2004}, wherein a relationship between the Mino time and coordinate time Fourier coefficients was derived for orbital functionals depending only on the radial and polar motions, i.e., $f=f(r, \theta)$. By defining 
\begin{equation}
\Delta{t}(\lambda)\equiv t(\lambda)-\Gamma\lambda,
\end{equation}
whose explicit form is given by Eqs.~3.24 and 3.31 in Ref.~\cite{Drasco_2004}, and 
\begin{align}
    \mathcal{F}(w^{r}, w^{\theta}, \omega)&= T\left[r(w^{r}), \theta(w^{\theta})\right]e^{i\omega\Delta{t}(w^{r},w^{\theta})} \nonumber \\
    &\phantom{=} \times f\left[r(w^{r}), \theta(w^{\theta})\right],\label{eq:mathcalFfunc}
\end{align}
the coordinate time Fourier coefficients $f_{k,n}$ of the expansion of $f(r,\theta)$ are given by~\cite{Drasco_2004}
\begin{align}
    f_{k,n}&=\frac{\mathcal{F}_{k,n}(\Omega_{k,n})}{\Gamma},\label{eq:DHCoeffs}
\end{align}where $\Gamma\equiv\Upsilon^{t}$ denotes the fundamental frequency of $t$ with respect to $\lambda$ and $\mathcal{F}_{k,n}$ are the Fourier coefficients of the Mino time expansion of $\mathcal{F}$, i.e.,
\begin{align}
    \mathcal{F}_{k,n}(\omega)&=\frac{1}{(2\pi)^{2}} \iint_{0}^{2\pi} \mathcal{F}(w^{r},w^{\theta}, \omega)e^{i(kw^{\theta}+nw^{r})}\!\ed w^{\theta}\ed w^{r}.\label{eq:mathcalFcoeffs}
\end{align}

Another approach to obtain the coordinate time Fourier coefficients was suggested in Ref.~\cite{Sopuerta_2011}. In this approach, one needs only to compute a time series of $f(r, \theta, \phi)$ and fit this to a truncated version of the coordinate time expansion for the coefficients $f_{m,k,n}$. In practice, we implement this approach by converting Eq.~\ref{eq:Kerr:BLExpansion} to an expansion in terms of sines and cosines via Euler's formula and converting the problem of fitting for $f_{m,k,n}$ into a linear system of equations which can be solved by use of any standard linear algebra package.

With the Fourier coefficients obtained using either approach, we can then estimate the derivatives of $f$ from Eq.~\ref{eq:Kerr:BLExpansion}~\cite{Sopuerta_2011}
\begin{align}
    \frac{\ed ^{N}f}{\ed t^{N}}&=\sum_{m, k, n}(-i\Omega_{m,k,n})^{N}f_{m,k,n}\,e^{-i\Omega_{m,k,n}t},\label{eq:Kerr:BLExpansionDeriv}
\end{align}with a similar expression holding for derivatives of $f$ with respect to $\lambda$ from the Mino time expansion.

Hereafter, we refer to the method of determining the Fourier coefficients via least-squares fitting as the ``Fourier fit'' approach and the method based on evaluating Eq.~\ref{eq:DHCoeffs} as the ``analytic'' approach, where it is understood that we numerically evaluate the analytic integrals which determine $f_{k, n}$.

Calculating the Fourier coefficients from Eq.~\ref{eq:DHCoeffs} is, in principle, the best approach because the derivation of these expressions in Ref.~\cite{Drasco_2004} is exact, utilizing the symmetries of the underlying geodesic orbit through the machinery of Hamilton--Jacobi theory. As such, this work is \emph{not} testing the correctness of this approach nor the Fourier expansion (Eq.~\ref{eq:Kerr:BLExpansion}). Rather, we are interested how the inevitable truncation of an orbital functional's Fourier expansion affects the accuracy of the time series reconstruction and time-derivative estimation using both methods, in addition to their respective computational efficiency which is important for applications of these approaches.

\section{High-order Time Derivatives of Geodesic Orbits}
\label{sec:BLDerivs}
In order to assess the accuracy of the time derivative estimation using the analytic and Fourier fit approaches described in the previous section, we must compute the ``true'' high-order coordinate time derivatives of the orbital functionals we consider. In this section, we prescribe a systematic procedure for computing these derivatives. First, we write down analytic Mino time derivatives of the geodesic orbit in a way that avoids: (i) excessively lengthy expressions that would be too computationally inefficient to be evaluated many times per time step in a numerical simulation, and (ii) the numerical problems which plague the first-order geodesic equations (as cast in the forms given by Eqs.~\ref{eq:PropTimeGeoODEs}, \ref{eq:MinoTimeGeoODEs}) at the poles $\dot{r}=0$, $\dot{\theta}=0$. Then, we describe a recursive procedure for transforming these to derivatives with respect to coordinate time.

One way to compute $\ed^{N}r/\ed{\lambda}^{N}$ is to take the square root of Eq.~\ref{eq:MinoTimeODEr} (with a positive sign as $r$ goes from periastron to apastron, and a negative sign for the reverse motion) and repeatedly take derivatives. However, the resulting expressions have at least two undesirable features. First, the expressions quickly get very lengthy as one takes higher-order derivatives due to the presence of a fractional power, making the numerical evaluations prone to truncation errors when evaluating them and symbolic calculations challenging. Second, the resulting expressions can misbehave at the radial and polar turning points, with numerical evaluation sometimes yielding complex numbers due to machine precision errors when approximating the argument of the square root as it approaches zero.

These two issues can be side-stepped by avoiding taking square roots in the first place. Differentiating both sides of Eq.~\ref{eq:MinoTimeODEr} with respect to $\lambda$ and rearranging, one obtains
\begin{align}
    \left(\frac{\ed^{2}{r}}{\ed\lambda^{2}}-\frac{1}{2}\frac{\ed{R}}{\ed{r}}\right)\frac{\ed{r}}{\ed\lambda}=0,\label{eq:RadialDerivChainRule}
\end{align}which holds for all $\lambda$, so that
\begin{align}
    \label{eq:d2r_d_lambda}
    \frac{\ed^{2}{r}}{\ed\lambda^{2}}&=\frac{1}{2}\frac{\ed{R}}{\ed{r}},
\end{align}where the corresponding expression for $\theta$ is obtained under the replacement $r\to\theta$ and $R\to\Theta$. There are no square roots present in Eq.~\ref{eq:d2r_d_lambda} and the functions $R(r)$ and $\Theta(\theta)$, defined in Eqs.~(\ref{eq:Rfunc}--\ref{eq:ThetaFunc}), contain positive integer powers and are thus more amenable to repeated differentiation. In particular, one can compute $\ed^{N}r/\ed{\lambda}^{N}$ by differentiating Eq.~\ref{eq:d2r_d_lambda} with respect to $\lambda$ and using the chain rule to express the resulting derivative in terms of derivatives of $R(r)$ and lower-order time derivatives of $r$, which will have already been computed at an earlier step. The computation of $\ed^{N}\theta/\ed{\lambda}^{N}$ follows similarly.

With the $\ed^N x^{\mu}/\ed \lambda^{N}$ at hand, the derivatives $\ed^N f/\ed \lambda^{N}$ can be computed for functions $f(\lambda)=f\left[r(\lambda), \theta(\lambda),\phi(\lambda)\right]$ explicitly expressed in terms of $r$, $\theta$ and $\phi$ by use of the chain rule. We can then compute $\ed^N f/\ed t^{N}$ by recursively using the chain rule, which allows one to express $\ed^N{f}/\ed{t}^{N}$ in terms of $\ed^{N}f/\ed{\lambda}^{N}$ and $\ed^{N}\lambda/\ed{t}^{N}$ (and lower order derivatives), only the latter of which remains to be computed. 

By simply writing
\begin{align}
    \frac{\ed\lambda}{\ed{t}}&=\frac{1}{\ed{t}/\ed\lambda},\label{eq:OneOverDtDλ}
\end{align}
we could proceed by replacing the denominator with $T(r,\theta)$ as per Eq.~\ref{eq:MinoTimeODEt} and directly applying the multivariate chain rule to take higher-order derivatives. However, continued differentiation in this fashion yields lengthy expressions as a result of the negative power of $T(r,\theta)$. A more efficient way to compute these derivatives is to write
\begin{subequations}
\begin{align}
    \frac{\ed^{2}\lambda}{\ed{t}^{2}}&=-\left(\frac{\ed{\lambda}}{\ed{t}}\right)^{3}\frac{\ed T}{\ed{\lambda}}.\label{eq:d2_lambda_dt_recursive}
\end{align}
\end{subequations}Instead of substituting Eq.~\ref{eq:OneOverDtDλ} directly, we can evaluate Eq.~\ref{eq:d2_lambda_dt_recursive} and its higher-order time derivatives recursively since the right-hand side will depend on lower-order derivatives of $\lambda$ with respect to $t$, which will have been computed in a previous step with the base case given by Eq.~\ref{eq:OneOverDtDλ}.

The procedure we have outlined can be used to convert the Mino time derivatives $\ed^{N} x^{\mu}/\ed\lambda^{N}$ to coordinate time derivatives $\ed^{N} x^{\mu}/\ed t^{N}$. In Sec.~\ref{ssec:SimpleExample}, we use these coordinate time derivatives of the orbit to compute the corresponding time derivatives of the test function $f=r\cos\theta$, which are used as the true values when computing fractional residuals of the estimated derivatives. We also use this procedure in Sec.~\ref{ssec:MultipoleExample} to exactly compute high-order coordinate time derivatives of the mass quadrupole expressed in harmonic coordinates. In this example, the Fourier fit is carried out in Mino time, so that from the analogous form of Eq.~\ref{eq:Kerr:BLExpansionDeriv}, one obtains estimates $\ed^{N}f/\ed \lambda^{N}$ which we convert to estimates with respect to coordinate time.

\section{Applications: Time Derivative Evaluation of Orbital Functionals}
\label{sec:Examples}

In the examples presented below, we compute the time series of a given orbital functional by numerically solving the first-order geodesic equations with respect to $\lambda$ using the \texttt{Julia} programming language package \texttt{DifferentialEquations.jl} with a fourth-order Runge-Kutta time integration and relative and absolute tolerances of $10^{-16}$. See, for example, Refs.~\cite{Drasco_2004,Sopuerta_2011} for further details regarding the numerical solution of these equations.

Unless otherwise stated, we choose orbital parameters $a=0.9$, $p=6$, $e=0.5$, and $x_{\mathrm{I}}=\sqrt{3}/2$, where the semi-latus rectum $p$, eccentricity $e$, and inclination parameter $x_{\mathrm{I}}$ are defined by
\begin{subequations}
    \begin{align}
        r&=\frac{p M}{1+e\cos{\chi_{r}}},\\
        x_{\mathrm{I}}&=\mathrm{sign}(L_{z})\sin{\theta_{\mathrm{min}}},
    \end{align}
\end{subequations}and $\chi_{r}$ is a monotonically increasing angle variable which is evolved in the numerical solution of the geodesic equations instead of $r$. 

We truncate the Fourier series expansion (Eq.~\ref{eq:Kerr:BLExpansion}) such that each index runs from $-\Nharm$ to $\Nharm$. When considering two and three fundamental frequencies, the total number of harmonic frequencies scales as $\Nharm^{2}$ and $\Nharm^{3}$, respectively. 

While the accuracy of the derivative estimation and the optimal choice of simulation parameters—such as integration time, the number of terms (harmonics) in the truncated expansion, and sampling rate—vary depending on the choice of orbital parameters, the overall qualitative behavior observed in the examples below remains broadly consistent across the entire parameter space (see, for instance, Tab.~\ref{tbl:FitResiduals}). As will be discussed below, more harmonics are especially required in the truncated Fourier series expansion as the eccentricity approaches unity, since, in this limit, expansion of $r$ converges increasingly slowly (see Ref.~\cite{Drasco_2004} and the references therein for further discussion of this slow convergence).

\subsection{A Simple Test Function}\label{ssec:SimpleExample}

\begin{figure*}
    \centering
    \includegraphics[width=\textwidth]{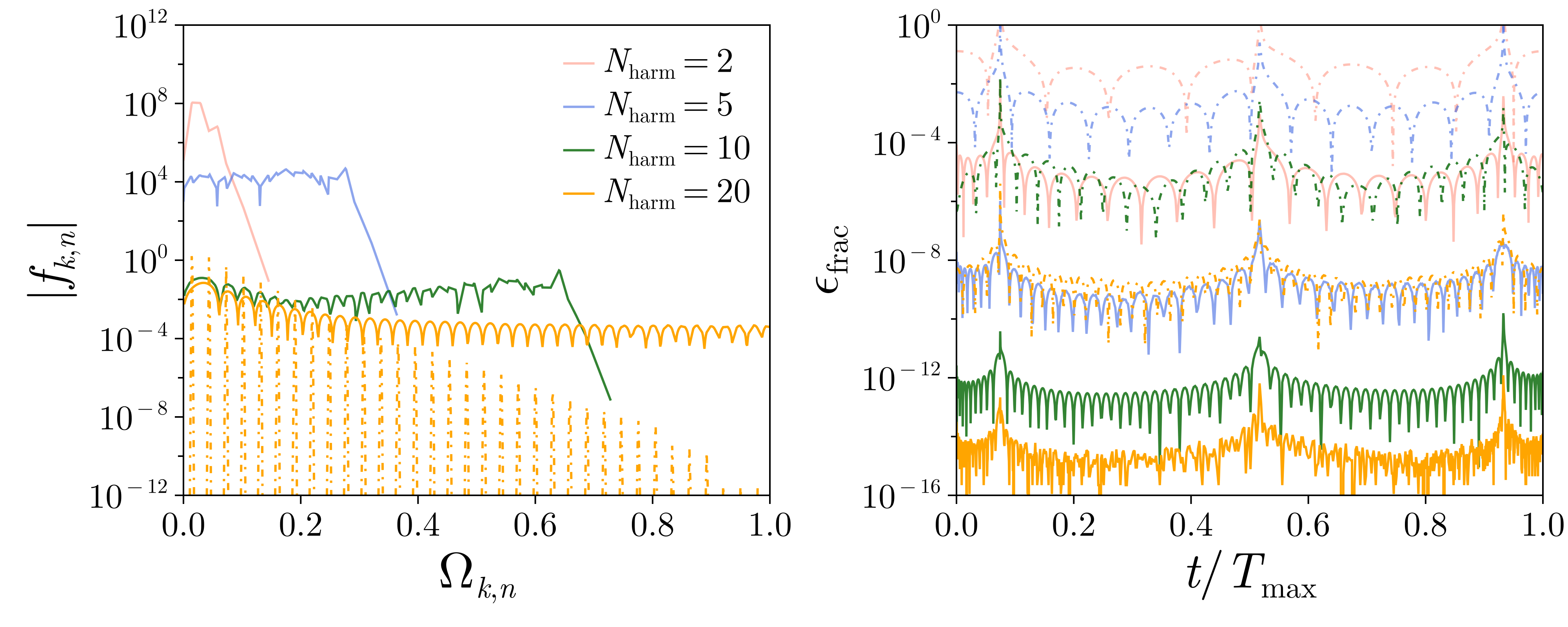}
    \caption{Reconstructed power spectrum of the orbital functional $f(t)=r\cos{\theta}$ \textbf{(left)} and fractional residual $\epsilon_{\mathrm{frac}}=|1-\tilde{f}(t)/f(t)|$ \textbf{(right)}, where $\tilde{f}(t)$ is the time series of $f$ reconstructed using the analytic \textbf{(dashed)} or Fourier fit \textbf{(solid)} methods. For clarity, we show only one dashed curve for $\Nharm=20$ in the left panel since the power spectra obtained from the analytic formula for the Fourier coefficients does not change pointwise as $\Nharm$ is increased, so all the curves for lower $\Nharm$ overlap. The underlying geodesic was sampled at a rate greater than twice the maximum frequency shown in the left panel, i.e., the Nyquist limit is greater than the upper $x$ limit. The fractional residual in the right panel spikes at some points as the orbital functional crosses zero and changes sign. As seen in the left panel, the power spectrum obtained using the fitting method changes with the number of terms included in the truncated expansion but does not converge to that obtained from the analytic method with increasing $\Nharm$. Nevertheless, the right panel shows that the fitting method provides time series reconstructions which are orders of magnitude more accurate than the analytic method for a given truncated expansion.
    }
    \label{fig:SimpleTestFunction:PowerResidualSubplot}
\end{figure*}

We first consider the orbital functional $f=r\cos{\theta}$, which was also studied as an example in Ref.~\cite{Drasco_2004}. Since the various integrals necessary to obtain the Fourier coefficients via  Eq.~\ref{eq:DHCoeffs} are performed over complete cycles of the radial and polar Mino time angle variables, we numerically evolve the underlying geodesic for a range of time $\Delta \lambda=1.01\,\Lambda_{\mathrm{max}}$, where $\Lambda_{\mathrm{max}}=2\pi / \min\{{\Upsilon_{r}, \Upsilon_{\theta}}\}$ to ensure that at least one cycle is executed with respect to both variables. In Fig.~\ref{fig:SimpleTestFunction:PowerResidualSubplot}, we show the power spectrum (left) and fractional residual (right) for different numbers of harmonics in the time series reconstruction of the test function $f$ using the analytic (dashed) and Fourier fit (solid) methods.

As shown in left panel of Fig.~\ref{fig:SimpleTestFunction:PowerResidualSubplot}, the power spectra obtained from the two reconstruction methods are distinct. The peaks in the power spectrum obtained from the analytic method decrease in power with increasing frequency, whereas the Fourier fit power spectrum is distributed approximately uniformly over a limited band of frequencies whose width increases with the number of harmonics. When $\Nharm$ is increased, the same underlying geodesic trajectory is used to compute the Fourier coefficients. In the analytic method, the integral (Eq.~\ref{eq:mathcalFcoeffs}) which determines the Fourier coefficients is independent for different frequencies, hence the power spectrum for a given number of harmonics is included in the power spectrum for a larger value of $\Nharm$. The Fourier fit power spectrum, however, drastically changes with $\Nharm$, despite each time fitting to the same time series array. The change in behavior of the power spectrum as $\Nharm$ is increased is not related to the number of points, $n_{p}$, in the time series relative to total the number of Fourier coefficients to be fit (which decreases with increasing $\Nharm$ since fitting to the same time series means $n_{p}$ is fixed). We have generically found that the fit does not depend greatly on the number of points, and here we have ensured that in each fit, $n_p$ is greater than the number of coefficients in the fit, so the linear system is overdetermined.

That the Fourier fit power spectrum depends on the value of $\Nharm$ suggests this approach overfits each truncated expansion to the time series array, as opposed to capturing the true harmonic structure inherited from the symmetries of the underlying geodesic orbit. Increasing the number of harmonics only expands the solution space for the linear system which determines the Fourier coefficients, allowing it to find increasingly better fits to the sampled time series $f(t)$. However, the corresponding power spectrum does not converge to that obtained from the analytic method.

The right panel of Fig.~\ref{fig:SimpleTestFunction:PowerResidualSubplot} illustrates that, although the Fourier fit approach fails to capture the correct harmonic structure in the power spectrum, the overfitting causes it to provide a much more accurate time series reconstruction than the analytic approach for a given value of $\Nharm$. Consistent with Ref.~\cite{Drasco_2004}, the analytic approach provides a reconstruction with a fractional residual $\fr\sim10^{-2}$ for $\Nharm=5$. The slow convergence of the Fourier series expansion (seen in the left panel of Fig.~\ref{fig:SimpleTestFunction:PowerResidualSubplot}) requires still larger values of $\Nharm$ to get a more accurate time series reconstruction necessary for accurate derivative estimation. In particular, with $\Nharm=20$ we obtain a reconstruction with $\fr\lesssim 10^{-8}$. In contrast, the overfitting in the Fourier fit approach appears to make it insensitive to the convergence rate of the orbital functional's true expansion. The fractional residual in the reconstruction obtained from this approach for $\Nharm=2$ and $\Nharm=5$ is already as accurate as that from the analytic method for $\Nharm=10$ and $\Nharm=20$, respectively. For $\Nharm=20$, the fractional residual decreases to $\fr\sim10^{-15}$.

\begin{figure}
    \centering
    \includegraphics[width=1.0\linewidth]{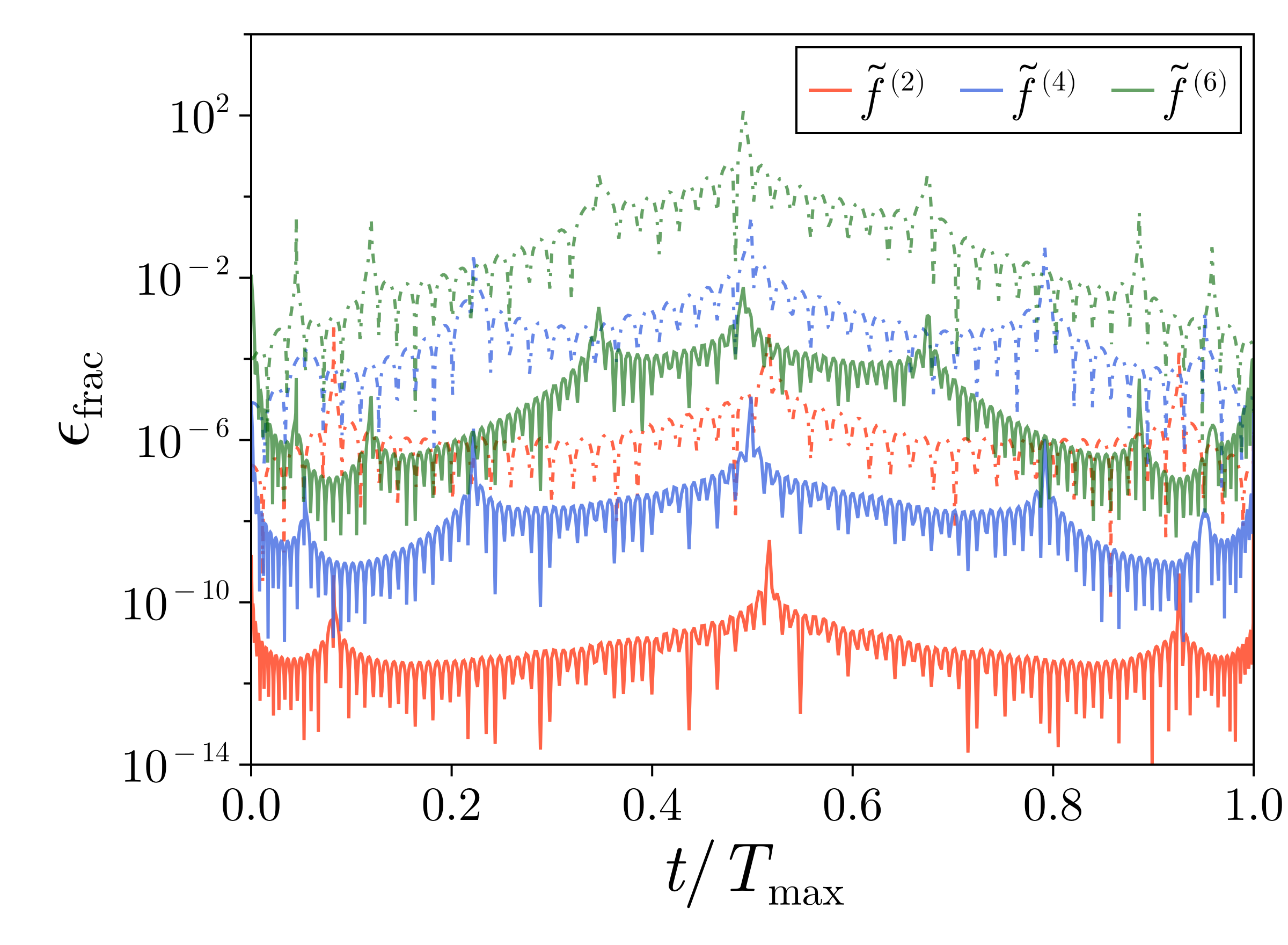}
    \caption{Derivative estimation fractional residual $\epsilon_{\mathrm{frac}}=|1-\tilde{f}^{(N)}(t)/f^{(N)}(t)|$, where $\tilde{f}^{(N)}(t)$ is the estimated $N$th (second, fourth, and sixth) time derivative obtained from the reconstructed time series of $f(t)=r\cos{\theta}$ from the analytic \textbf{(dashed)} or Fourier fit \textbf{(solid)} methods. The underlying Fourier coefficients used to evaluate Eq.~\ref{eq:Kerr:BLExpansionDeriv} are the same as those depicted in Fig.~\ref{fig:SimpleTestFunction:PowerResidualSubplot} for $\Nharm=20$. Although the fitting method fails to correctly capture the harmonic structure of $f$ due to overfitting, the resulting increased accuracy in the reconstructed $\tilde{f}$ relative to the analytic method yields derivative estimations which are more accurate than the analytic method by several orders of magnitude for each derivative order.
    }
    \label{fig:SimpleTestFunction:DerivativeEstimation}
\end{figure}

The importance of obtaining an accurate time series reconstruction for derivative estimation can be seen in Fig.~\ref{fig:SimpleTestFunction:DerivativeEstimation}, where $\epsilon_{\mathrm{frac}}$ in the estimated second, fourth, and sixth time derivatives of $f(t)=r\cos{\theta}$ using each reconstruction method is shown. The fractional residual in a given estimated derivative is always several orders of magnitude less in the Fourier fit compared to the analytic approach, while $\fr$ increases more rapidly for each increase in derivative order in the former compared to the latter method. Note that one expects some deterioration in accuracy as one tries to estimate higher-order derivatives because the power in larger harmonic frequencies neglected in a given truncation grows more rapidly with derivative order than those included in the truncation (see Eq.~\ref{eq:Kerr:BLExpansionDeriv}). That the deterioration is more rapid in the Fourier fit approach is likely also a consequence of overfitting and failing to get the correct power spectrum, since only with the correct Fourier coefficients does the infinite expansion (Eq.~\ref{eq:Kerr:BLExpansion}) contain complete information both about $f$ and its time derivatives. In both cases, this phenomenon means increasingly accurate fits of the time series array of $f$ are required to estimate higher and higher-order derivatives to within a fixed value of $\fr$.

\begin{table}[hbt!]
\begin{tabular}{cccccc}
\cline{3-6}
\cline{4-6}
\cline{5-6}
\cline{6-6}
\multicolumn{1}{c}{} & \multicolumn{1}{c|}{} & \multicolumn{2}{c|}{Analytic} & \multicolumn{2}{c|}{Fourier Fit} \tabularnewline
\hline
\multicolumn{1}{|c|}{\shortstack{$e$}} & \multicolumn{1}{c|}{\shortstack{$x_{\mathrm{I}}$}} & \multicolumn{1}{c|}{\shortstack{$\tilde{f}$}} & \multicolumn{1}{c|}{\shortstack{$\tilde{f}^{(6)}$}} & \multicolumn{1}{c|}{\shortstack{$\tilde{f}$}} & \multicolumn{1}{c|}{\shortstack{$\tilde{f}^{(6)}$}} \tabularnewline
\hline
\hline
$0.1$ & $0.97$ & $10^{-11}$ & $10^{-4}$ & $10^{-15}$ & $10^{-5}$ \tabularnewline
$0.1$ & $0.26$ & $10^{-11}$ & $10^{-5}$ & $10^{-15}$ & $10^{-6}$ \tabularnewline
$0.8$ & $0.97$ & $10^{-4}$ & $10^{1}$ & $10^{-10}$ & $10^{-3}$ \tabularnewline
$0.8$ & $0.26$ & $10^{-4}$ & $10^{1}$ & $10^{-9}$ & $10^{-3}$ \tabularnewline
\hline
\hline
\end{tabular}
\caption{Order of magnitude of the fractional residual in the reconstruction of $f=r\cos{\theta}$ and its estimated sixth coordinate time derivative for various combinations of eccentricities and inclinations, with the remaining orbital parameters otherwise the same as the examples presented in Figs.~\ref{fig:SimpleTestFunction:PowerResidualSubplot}--\ref{fig:SimpleTestFunction:DerivativeEstimation}. For the larger eccentricity cases, we quote $\fr$ for $\tilde{f}^{(6)}$ near the edges since the the true value is close to zero near the center of the time series array. We have fixed $\Nharm=20$ to compare how the error changes with the eccentricity and inclination in each method. As expected, $\fr$ significantly increases with eccentricity, since $r$ converges increasingly slowly for larger values of $e$~\cite{Drasco_2004}. One can obtain better fits than presented in this table for $e=0.8$ by increasing the number of harmonics. For instance, for $\Nharm=40$ we obtain fraction residuals in $\tilde{f}$ and $\tilde{f}^{(6)}$ of $10^{-5}$ and $10^{-1}$ for the analytic method, respectively, and $10^{-15}$ and $10^{-6}$ for the Fourier fit method.}
\label{tbl:FitResiduals}
\end{table}

To summarize, in this example, fitting the time series of the orbital functional $f(t)=r\cos{\theta}$ to a truncation of its Fourier series expansion is \emph{not} a reliable approach for accurately extracting the Fourier coefficients due to overfitting. However, if one is only interested in an accurate reconstruction of the time series and its time derivatives, this method works well, having a fractional residual less than the analytic approach by several orders of magnitude. For both methods, the derivative estimation worsens by a couple of orders of magnitude for each increase in the derivative order. 
In Tab.~\ref{tbl:FitResiduals} we present the fractional residuals from similar applications of each method to orbits with various combinations of small and large eccentricities and inclinations. Although, with a fixed number of harmonics, the overall accuracy of the reconstructions decrease for the more eccentric orbits, the fitting method is less sensitive than the analytic method to the slowing convergence of the orbital functional's Fourier series expansion. The examples presented in Table~\ref{tbl:FitResiduals} demonstrate that the overall qualitative behavior of estimating these high-order time derivatives remains consistent across a broad range of parameters. It is important to note that, depending on the radial and angular structure of the orbital functional, more extreme orbits will require a larger number of harmonics to estimate derivatives with a fixed accuracy. For example, for the Fourier fit method applied to the $e=0.8$ cases in Tab.~\ref{tbl:FitResiduals} to provide estimates with a similar accuracy to the $e=0.1$ cases, the number of harmonics must be doubled from $\Nharm=20$ to $\Nharm=40$.

The Fourier fit approach has the additional advantage of being at least as computationally efficient as the analytic method when solving a linear system which either has a unique solution or is overdetermined. The additional computational cost in the analytic method is set by the convergence of the Fourier expansion of $\Delta{t}(w^{r}, w^{\theta})$, which appears in Eq.~\ref{eq:mathcalFfunc}. For a fit with $n$ points in the sampled time series array of $f$ and $m$ nonzero harmonic frequencies, Fourier fitting has a complexity of $\mathcal{O}(m^3+m^2 n)$ floating point operations compared to $\mathcal{O}(Bmn^2)$ for the analytic method, where $B$ is the total number of radial and polar harmonics kept in the truncated expansion of $\Delta {t}(w^{r}, w^{\theta})$. In the case $n\gg m$, the linear dependence of the Fourier fitting's complexity on $n$ is dominated by the quadratic dependence for the analytic method; for $n\sim m$ the complexities only differ by the additional factor of $B$ for the analytic method. The chosen value of $B$ will vary depending on the orbital parameters, and, in particular, a larger number of radial harmonics will be required as the eccentricity approaches unity due to the slowing convergence of the radial coordinate's Fourier expansion. For the fits shown in Figs.~\ref{fig:SimpleTestFunction:PowerResidualSubplot} and \ref{fig:SimpleTestFunction:DerivativeEstimation}, we kept $10$ and $30$ polar and radial harmonics, respectively, in the expansion of $\Delta{t}(w^{r}, w^{\theta})$, giving $B=40$.

We have empirically found more accurate coordinate time derivative estimation by use of an alternative approach of fitting the time series to its Fourier series expansion in Mino time and converting the estimated derivatives with respect to Mino time into coordinate time derivatives. (The procedure for carrying out this conversion was described in the last paragraph in Sec.~\ref{sec:BLDerivs}.) In the specific case shown in Figs.~\ref{fig:SimpleTestFunction:PowerResidualSubplot} and~\ref{fig:SimpleTestFunction:DerivativeEstimation}, the Mino time fits yield a very similar power spectrum to the coordinate time fits, while $\fr\sim10^{-6}$ for the sixth derivative at the center, an improvement by two orders of magnitude compared to coordinate time fits. 

Since we are interested in the accurate time derivative estimation and not power spectrum reconstruction, we focus in the next section on an efficient application of the more accurate Mino time Fourier fit approach to a set of more complicated orbital functionals which additionally depend on the azimuthal coordinate $\phi$. The analytic method applies to orbital functions of the form $f=f(r,\theta)$, so we restrict our attention to only the Mino time Fourier fit method.

\subsection{Mass and Current Multipoles in Harmonic Coordinates}
\label{ssec:MultipoleExample}

We now turn to the application of the Mino time Fourier fit approach to estimating the time derivatives of a set of orbital functionals with a more complex dependence on the underlying geodesic orbit. In particular, we consider as an example the mass and current multipoles required in the so-called ``\texttt{Chimera}'' EMRI kludge scheme introduced in Ref.~\cite{Sopuerta_2011}. We first describe in Sec.~\ref{sssec:MultipoleDerivsAnalytic} the computation of the time derivatives of these multipoles from analytic expressions, which we compare in Sec.~\ref{sssec:MultipoleDerivEstimation} to the estimates provided by the Fourier fitting scheme.

We denote by $M$ the mass of the primary (supermassive) black hole in the EMRI system and by $m$ the mass of the secondary (stellar mass) black hole, where the mass ratio $q=m/M\ll 1 $. We also define $m_{\tot}\equiv M+m$, $\delta m\equiv M-m$, and the symmetric mass ratio $\eta\equiv Mm/m_{\tot}^{2}=q\,(1+q)^{-2}$. For the fits shown in Fig.~\ref{fig:MultipoleMoments:DerivativeEstimation} we take $q=10^{-5}$.

\subsubsection{Analytic Time Derivatives}\label{sssec:MultipoleDerivsAnalytic}

We consider mass and current quadrupoles which are expressed in terms of the location of the EMRI system's secondary object, whose trajectory, in the so-called ``osculating orbits'' approach, is treated as geodesic on orbital time scales, where $T_{\mathrm{orbit}}/T_{RR}\sim q\ll 1$. In the \texttt{Chimera} kludge scheme, the inspiralling worldline of the secondary object is then built as a sequence of geodesics between which the constants of motion are updated using mass and current quadrupoles obtained from multipolar post-Minkowskian and post-Newtonian  expansions in terms of harmonic coordinates  $\{x_{\mathrm{H}}^{i}\}$~\cite{Sopuerta_2011, Blanchet_2010, Arun_2008},
\begin{subequations}
\begin{align}    M_{ij}&=\eta\,m_{\tot}\,x_{\langle ij\rangle},\label{eq:Mass_quad}\\
    S_{ij} &=-\eta\,\delta{m}\,\epsilon_{ab\langle i}\,x_{j \rangle}{}^{a}\,\dot{x}^{b},\label{eq:Current_quad}
\end{align}\label{eq:MultipoleMoments}
\end{subequations}where we have used multi-index notation for the spatial (Latin) indices, raising/lowering in Eq.~\ref{eq:MultipoleMoments} is done with respect to the Kronecker delta symbol, and the symmetric-trace-free projections are given by
\begin{subequations}
\begin{align}
    x^{\langle ij \rangle} &= x^{ij} - \frac{1}{3}(x^{l}x_{l})\delta^{ij},\label{eq:STF xij}\\
    \epsilon_{kl\langle i}x_{j\rangle}&=\frac{1}{2}\left(\epsilon_{kli}x_{j}+\epsilon_{klj}x_{i} \right)\nonumber\\
    &\quad- \frac{1}{3}\delta_{ij}\left(\epsilon_{kl1}x_{1} + \epsilon_{kl2}x_{2} + \epsilon_{kl3}x_{3}\right)\label{eq:STF εkli * xj}.
\end{align}
\end{subequations}

We adopt the same harmonic coordinate system as in Ref.~\cite{Sopuerta_2011}, whose transformations from Boyer--Lindquist coordinates are given by
\begin{subequations}
\begin{gather}
    \label{eq:t_Harm_forward}
    t_\HH = t,\\
    \label{eq:x_Harm_forward}
    x_\HH = \rho\,\sin{\theta}\cos{\xi},\\
    \label{eq:y_Harm_forward}
    y_\HH = \rho\,\sin{\theta}\sin{\xi},\\
    \label{eq:z_Harm_forward}
    z_\HH = \left(r-M\right)\cos{\theta},
\end{gather}\label{eq:HarmonicCoords}
\end{subequations}where
\begin{subequations}
\begin{align}
    \label{eq:rho_def}
    \rho &\equiv \sqrt{\left(r-M\right)^{2}+a^{2}},\\
    \label{eq:xi_def}
    \xi &\equiv \phi - \Phi(r),
\end{align}
\end{subequations}and
\begin{subequations}
\begin{align}
    \label{eq:Phi_of_r_def}
    \Phi(r) &= \frac{\pi}{2}-\atan{\frac{r-M}{a}}-\frac{a}{2\sqrt{M^{2}-a^{2}}}\ln{\left(\frac{r-r_{-}}{r-r_{+}}\right)}.
\end{align}
\end{subequations}

The analytic time derivatives of the mass and current quadrupoles can be computed  by differentiating Eq.~\ref{eq:MultipoleMoments} with respect to coordinate time to obtain expressions in terms of time derivatives of the harmonic coordinates, which, in turn, can be computed by differentiating Eq.~\ref{eq:HarmonicCoords} and using the procedure described in Sec.~\ref{sec:BLDerivs} to compute the high-order coordinate time derivatives of the underlying geodesic trajectory.

\subsubsection{Time Derivative Approximation}
\label{sssec:MultipoleDerivEstimation}

In the original implementation of the \texttt{Chimera} kludge scheme, Jacobians and Hessians are used to transform up to second-order time derivatives of the geodesic orbit described in Boyer--Lindquist coordinates into time derivatives of the harmonic coordinates. With this approach, up to the second time derivative of $M_{ij}$ and the first derivative of $S_{ij}$ can be computed analytically. The expressions for the self-force in the \texttt{Chimera} include up to the eighth time derivative of $M_{ij}$, thus requiring six additional derivatives of $\ddot{M}_{ij}$ (where the overdot denotes differentiation with respect to $t$) to be estimated. In this section, we consider the accuracy of this additional derivative estimation by computing $\ddot{M}_{ij}$ analytically and taking as the test orbital functional $f=\ddot{M}_{12}$.

\begin{figure*}
    \centering
    \includegraphics[width=1.0\linewidth]{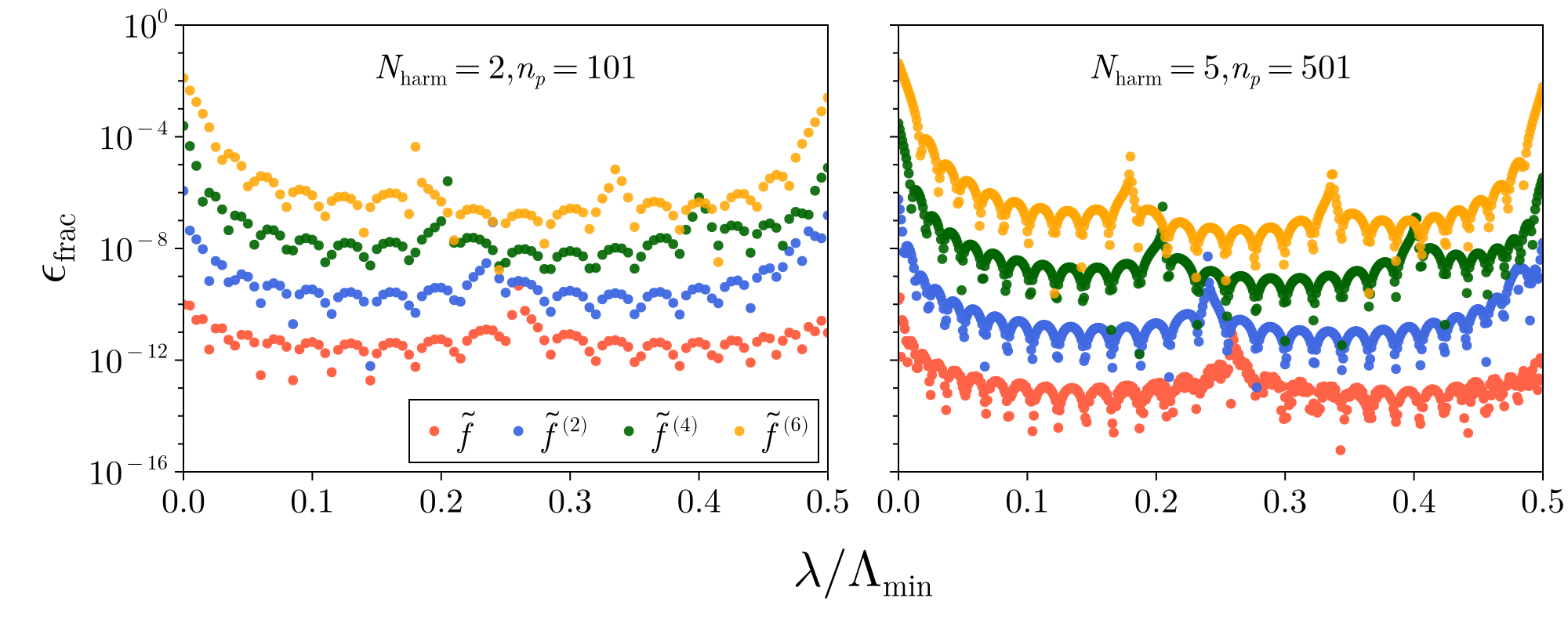}
    \caption{Reconstruction and derivative estimation fractional residuals $\epsilon_{\mathrm{frac}}=|1-\tilde{f}^{(N)}(t)/f^{(N)}(t)|$, where $\tilde{f}^{(N)}(t)$ is the estimated $N$th (zeroth, second, fourth, and sixth) time derivative of $f(t)=\ddot{M}_{12}$, a mass quadrupole component. These residuals are obtained by fitting a truncated Mino time expansion of $f$ for its $\lambda$-coefficients, evaluating the Mino time version of Eq.~\ref{eq:Kerr:BLExpansionDeriv}, and then converting these to derivatives with respect to coordinate time. \textbf{Left:} a fit with $\Nharm=2$ and $n_{p}=101$ which has $\epsilon_{\mathrm{frac}}\sim 10^{-6}$ near the center of the array for the estimated $\tilde{f}^{(6)}$ (i.e., the eighth coordinate time derivative of $M_{12}$). \textbf{Right:} a fit with $\Nharm=5$ and $n_{p}=501$ which has $\epsilon_{\mathrm{frac}}\sim10^{-7}$ near the center for the estimated $\tilde{f}^{(6)}$. In this example, the underlying geodesic was evolved for a time $\Delta\lambda=0.5\Lambda_{\mathrm{min}}$, where $\Lambda_{\mathrm{min}}=2\pi/\max\{\Upsilon_{r}, \Upsilon_{\theta}, \Upsilon_{\phi}\}$. Though the fit shown in the right panel has many more terms in the truncated Fourier expansion and a higher sampling rate of $f$, the fit in the left panel is less accurate by only roughly an order of magnitude, while being computationally faster by a factor about two orders of magnitude with respect to the case shown in the right panel.}
    \label{fig:MultipoleMoments:DerivativeEstimation}
\end{figure*}

In Fig.~\ref{fig:MultipoleMoments:DerivativeEstimation} we show the fractional residual of the time series reconstruction of $f(t)=\ddot{M}_{12}$, and its derivatives with respect to coordinate time up to sixth order (i.e., up to $\ed ^{8}M_{12}/\ed t^8$) using the Mino time Fourier fit method. Shown in the left panel of Fig.~\ref{fig:MultipoleMoments:DerivativeEstimation} is a fit performed to a time series array of $f$ containing $n_{p}=101$ points and using $\Nharm=2$ harmonics in the fit. On the right, a fit with $n_{p}=501$ and $\Nharm=5$ is shown. The fit with $\Nharm=2$ performs much better than that shown in Figs.~\ref{fig:SimpleTestFunction:PowerResidualSubplot} and \ref{fig:SimpleTestFunction:DerivativeEstimation} because we now consider all three fundamental frequencies for which the total number of harmonic frequencies used in the fit scales like $\Nharm^3$ instead of $\Nharm^2$ for two fundamental frequencies. 

Despite the increased complexity of the underlying orbital functional and the additional dependence in this case on $\phi$, the fits shown in Fig.~\ref{fig:MultipoleMoments:DerivativeEstimation} are still very accurate: for the fit with $\Nharm=5$, the fractional residual $\epsilon_{\mathrm{frac}}\sim10^{-13}$ in the time series reconstruction of $\ddot{M}_{12}$ and, in its estimated sixth coordinate time derivative, $\epsilon_{\mathrm{frac}}\sim10^{-7}$ near the center. Remarkably, the fit with fewer harmonics and a smaller sampling rate of $\ddot{M}_{12}$ is only roughly an order of magnitude worse in the fractional residual for each derivative order. In absolute terms, the residual remains very low for the fit with $\Nharm=2$ ($\epsilon_{\mathrm{frac}}\sim 10^{-6}$ in the sixth derivative at the center) and it is computed $\sim 100$ times faster.

In the previous example, we remarked that the fitting approach was always at least as efficient as the analytic method for the same number of points and harmonics. An additional feature of the Fourier fit method illustrated in this example is the freedom in this approach to reduce the computational cost by fitting with fewer points in the time series array. In the analytic approach, one has to evolve the underlying geodesic for at least one orbital cycle in each coordinate in order to evaluate the various integrals necessary to compute the Fourier coefficients. The number of points in this case is limited by having a sufficiently small step size such that the truncation error in numerically evaluating the integrals does not corrupt the high-frequency Fourier coefficients, whose importance increases as the expansion of $f$ converges more slowly. However, the Fourier fit approach does not face the same restrictions. In principle, we are free to shorten the time range so that, with a fixed tolerance in the size of the fractional residual: (i) fewer harmonics are necessary in the truncated Fourier series expansion and (ii) fewer points are needed in the orbital functional time series array. For the orbital functionals we have considered (simple test functions which combine $r$ and trigonometric functions of $\theta$ and $\phi$), we have found that the Mino time fits perform best on time scales $0.1 \leq \Delta\lambda / \Lambda_{\mathrm{min}} \leq 1.0 $ with $n_{p}<1000$, although the time scales might vary among different classes of orbital functionals and in more extreme regions in the orbit parameter space. For larger intervals of time and a fixed tolerance in the fractional residual, one needs more harmonics and points in the time series array which increases the computational cost. 

    The behavior of the derivative estimation described in this section and shown in Fig.~\ref{fig:MultipoleMoments:DerivativeEstimation} for $f=\ddot{M}_{12}$ is similar for the other independent components as well as for the current quadrupole and the mass octupole. Since these are the three multipole moments used for computing the fluxes in the original version of \texttt{Chimera} kludge scheme, the example presented in this section both quantifies the magnitude of the errors incurred in computing the fluxes by approximating these high-order derivatives as well as provides a scheme for computing them from analytic expressions. Although in this example we fitted components of the second time derivative of $M_{ij}$, the behavior of the fits to $f=M_{ij}$ is broadly similar to that shown in Fig.~\ref{fig:MultipoleMoments:DerivativeEstimation}, suggesting that already with $\Nharm=2$, the Mino time Fourier fit approach can be used to estimate the mass quadrupole contribution to the gravitational waveform (truncated at octupolar order $l=3$)~\cite{Sopuerta_2011, Blanchet_2002}
\begin{align}
    h_{ij}^{\TT}&=\frac{2}{R}\,\ddot{M}_{ij}+\frac{2}{3R}\dddot{M}_{ijk} n^{k}+\frac{8}{3R}\,\epsilon^{kl}{}_{(i}\,\ddot{S}_{j)k}n_{l}\nonumber\\
    &\quad+\frac{1}{R}\,\epsilon^{kl}{}_{(i}\,\dddot{S}_{j)km}\,n^{l}\,n^{m},\label{eq:hij}
\end{align}with fractional residuals on the order of $10^{-8}$. These methods become particularly useful for incorporating the higher $l$ contributions in Eq.~\eqref{eq:hij} which contain higher-order time derivatives.

\section{Discussion}
\label{sec:Discussion}

We have compared two methods for estimating high-order coordinate time derivatives of Kerr orbital functionals. The first method consists of determining the Fourier coefficients in the coordinate-time expansion of generic functions of the form $f(t)=f[r(t),\theta(t)]$ by numerically evaluating the closed-form analytic formulae derived in Ref.~\cite{Drasco_2004}. In the second approach, implemented in Ref.~\cite{Sopuerta_2011}, one fits a time series array of $f(t)$ to a truncated version of its Fourier series expansion to determine the Fourier coefficients. We performed this comparison on simple test functions and for more complex orbital functionals of mass and current quadrupoles expressed in harmonic coordinates.

In the case of a simple orbital functional $f=r\cos{\theta}$, the Fourier fit overfits the time series, causing the power spectrum to markedly differ from that obtained using the analytic method, instead preferring to uniformly distribute power over a limited range of frequencies. However, this overfitting, for an appropriate choice of fitting parameters (e.g., the time range, number of points, and number of harmonics), allows for a time series reconstruction and derivative estimation that is several orders of magnitude more accurate than the analytic approach. A similar level of accuracy is also obtained when applying the fitting method to the more complicated case of black hole mass and current multipoles. 

The Fourier fit method not only offers more accurate reconstructions and derivative estimations, but it is also more computationally efficient. This efficiency arises from both the inherent computational complexity of each method and the flexibility of the fitting scheme which allows for a shorter underlying geodesic time range. As a result, it requires fewer points and harmonics in the fitting process.

We have also developed a hybrid version of the Fourier fit approach in which one instead fits the orbital functional $f$ to its Fourier series expansion in Mino time, estimates the derivatives of $f$ with respect to $\lambda$, and then converts these to derivatives with respect to coordinate time. The last step in this approach relies on computing high-order coordinate time derivatives of the underlying geodesic orbit in Boyer--Lindquist coordinates, for which we have outlined an efficient procedure based on analytic expressions for high-order derivatives of the geodesic motion with respect to $\lambda$. For orbital functionals with closed-form expressions in terms of $\{r, \theta,\phi\}$, this procedure also offers the option of efficiently computing high-order derivatives exactly (to within machine precision), as demonstrated in Sec.~\ref{sssec:MultipoleDerivsAnalytic}.

In cases where one cannot write $f$ explicitly in terms of Boyer--Lindquist coordinates of the orbit, our examples demonstrated that the proposed hybrid Mino time Fourier fit approach provides an efficient estimation of high-order time derivatives. In particular, at the center of the sampled time series array, this approach estimates the sixth coordinate time derivative of $f=\ddot{M}_{12}$ (the eight time derivative of the mass quadrupole component $M_{12}$) with a fractional residual $\fr\sim 10^{-6}$.

The methods developed in this work provide a general framework for efficiently and accurately evaluating high-order time derivatives along Kerr worldlines which can be treated as geodesic on short time scales. As a concrete application, we considered derivative estimation for the mass and current multipoles required by the EMRI kludge scheme introduced in Ref.~\cite{Sopuerta_2011} to locally approximate the dissipative part of the gravitational self-force and to construct the gravitational waveform, though our methods are equally applicable to a broader range of problems in relativistic astrophysics. We leave to a future work a new implementation of the \texttt{Chimera} kludge scheme with these derivative estimation methods and a systematic comparison with alternative EMRI waveform models.

Our results provide a way to assess the errors that EMRI kludge schemes can incur by approximating the necessary high-order time derivatives when including high-order multipoles in the computation of radiation-reaction fluxes and waveforms. The examples considered also explicitly show how the error in the derivative estimation can deteriorate by several orders of magnitude for each increase in derivative order, suggesting the methods as implemented here have a limited range of applicability when attempting to include higher-order multipoles. The procedure we describe for computing these derivatives directly from the equations of motion thus becomes especially useful for the inclusion of higher-order post-Newtonian terms, as well as being far less computationally expensive than the other two methods, which consist of evaluating $2$D integrals numerically or solving a least-squares problem.

Although the methods explored in this work are, in principle, applicable to a broad class of orbital functionals, the number of harmonics required, and therefore the computational cost of applying these methods, is highly dependent on the specific orbital functional under consideration and the orbital parameters of the underlying trajectory. In the $f=r\cos{\theta}$ example, increasing the eccentricity of the underlying orbit from $e=0.1$ to $e=0.8$ required double the number of harmonics in the Fourier fit method to estimate time derivatives with the same accuracy (see Tab.~\ref{tbl:FitResiduals}). These methods should apply to orbits with more extreme eccentricities, but may require an increasing number of harmonics in the fit, depending on the orbital functional. We leave to future work a systematic study of how the error incurred from the Fourier fit method behaves as a function of the number of harmonics for a wider class of orbital functionals with different radial and angular structures.
    
Several directions remain for refining the methods developed in this work. In particular, given an orbital functional $f$ for which one can analytically compute the first (and possibly higher-order) time derivative(s), one may be able to better constrain the Fourier coefficients by \emph{simultaneously} fitting $f$ to its truncated expansion and $\dot{f}$ to the first time derivative of the same expansion. Additionally, we have here only studied analytic methods, but it is possible that other techniques, such as automatic differentiation (see, e.g., Refs.~\cite{2015arXiv150205767G, Edwards:2023sak}), might be suitable for estimating high-order time derivatives. This is particularly of interest for applications to EMRIs since automatic differentiation could also be applied to the problem of parameter estimation from catalogs of gravitational waveforms.

\acknowledgments

We are grateful to D.~Cutler, E.~Levati, F.~Pretorius, L.~Stein, N.~Yunes, H.~Zhu and especially S.~Hughes and C.~Sopuerta for clarifying conversations and comments.

\bibliographystyle{apsrev4-1}
\bibliography{refs}
\end{document}